\documentclass[12pt]{article}

\RequirePackage{amsmath, amsthm, amsfonts, graphicx, bm, natbib}
\RequirePackage{setspace, microtype, dsfont, subcaption, appendix}
\RequirePackage{multirow, colortbl, subcaption}
\RequirePackage{hyperref}
\RequirePackage[raggedright]{titlesec}
\RequirePackage{enumitem, xcolor}

\DisableLigatures[f]{encoding=*, family=*}
\numberwithin{equation}{section}
\numberwithin{figure}{section}
\numberwithin{table}{section}

\def\by{\bm{y}}
\def\bx{\bm{x}}

\def\eps{\epsilon}

\def\g{\gamma}

\def\a{\alpha}
\def\b{\beta}

\def\bmu{\boldsymbol \mu}

\def\bb{\boldsymbol \beta}
\def\bxi{\bm{\xi}}
\def\Sig{\boldsymbol \Sigma}

\def\N{\mathcal{N}}

\makeatletter
\let\@fnsymbol\@arabic
\makeatother

\title{\bf Modelling age-related changes in executive functions of soccer players}
\author{Vincent Chin\footnote{Australian Research Council Industrial Transformation Training Centre for Data Analytics for Resources and Environments, The University of Sydney, Sydney, Australia.} \textsuperscript{,}\footnote{School of Mathematics and Statistics, The University of Sydney, Sydney, Australia.} , 
Adam Beavan\footnote{TSG ResearchLab gGmbH, Zuzenhausen, Germany} , 
Job Fransen\footnote{Human Performance Research Centre, School of Sport, Exercise and Rehabilitation, University of Technology, Sydney, Australia.} ,
Jan Mayer\footnote{
Performance Diagnostics, TSG 1899 Hoffenheim Fu{\ss}ball-Spielbetriebs GmbH, Zuzenhausen, Germany.} , \\
Robert Kohn\footnotemark[1] \textsuperscript{,}\footnote{Australian Research Council Centre of Excellence for Mathematical \& Statistical Frontiers, The University of Melbourne, Victoria, Australia.} \textsuperscript{,}\footnote{School of Economics, Australian School of Business, University of New South Wales, Sydney, Australia.} ,
Louise M. Ryan\footnotemark[6] \textsuperscript{,}\footnote{School of Mathematical and Physical Sciences, University of Technology Sydney, Sydney, Australia.} \textsuperscript{,}\footnote{Harvard T.H. Chan School of Public Health, Harvard University, Cambridge, USA.} \thinspace \thinspace and
Scott A. Sisson\footnotemark[6] \textsuperscript{,}\footnote{UNSW Data Science Hub \& School of Mathematics and Statistics, University of New South Wales, Sydney, Australia.}
}
\date{}

\begin{document}
\maketitle

\begin{abstract}
\onehalfspacing
The widespread popularity of soccer across the globe has turned it into a multi-billion dollar industry. As a result, most professional clubs actively engage in talent identification and development programmes.  Contemporary research has generally supported the use of executive functions -- a class of neuropsychological processes responsible for cognitive behaviours -- in predicting a soccer player's future success. However, studies on the developmental evolution of executive functions have yielded differing results in their structural form (such as inverted U-shapes, or otherwise).  This article presents the first analysis of changes in the domain-generic and domain-specific executive functions based on longitudinal data measured on elite German soccer players. Results obtained from a latent variable model show that these executive functions experience noticeable growth from late childhood until pre-adolescence, but remain fairly stable in later growth stages. As a consequence, our results suggest that the role of executive functions in facilitating talent identification may have been overly emphasised. 
\end{abstract}
\begin{flushleft}
\end{flushleft}
\doublespacing

\newpage
\section{Introduction}

Sports activity has long been an integral part of everyday life, and is synonymous with a healthy lifestyle due to well-established relationships between an adequate level of physical exercise and physiological benefits such as reduced risk of developing coronary heart disease \citep{bassuk2005epidemiological, sofi2008physical} and improved cognitive performance \citep{kramer2007capitalizing, hillman2008smart}.  Soccer
(also known as football) is one of the most popular sports in the world, which is evident from the over 3.5 billion total viewerships \citep{fifa2018fifa} it garnered for the quadrennial World Cup tournament in 2018.  Due to its ubiquitous global presence, the sport has grown into a multi-billion dollar industry over the last few decades, with soccer clubs investing heavily in talent identification programmes \textcolor{black}{including those involving measures of executive functioning. Formally, executive functions are cognitive processes that facilitate decision-making \citep{best2010developmental, furley2016working} and goal-oriented behaviours \citep{zelazo2004executive} based on information within the context of the task \citep{alvarez2006executive}. Some examples of these processes are problem-solving, sustained attention, resistance to interference, utilisation of feedback, and multi-tasking \citep{grafman1999importance,burgess2000cognitive,chan2008assessment}.}

\textcolor{black}{The proposition advocating for using executive functions testing within the talent identification process is based on the considerable amount of evidence indicating that neuropsychological evaluations provide a useful indicator for performance success in young soccer players. For example, \cite{verburgh2014executive} revealed that elite players show superior motor inhibition as measured by a stop signal task compared to amateur players of the same age, and this finding is reaffirmed in other similar experiments \citep{huijgen2015cognitive, sakamoto2018possible}. However, longitudinal studies of the developmental trajectories of executive functions across different stages of life in an athlete population are lacking in the literature as previous examinations focus on a general population \citep{zelazo2004executive,huizinga2010age, zelazo2012hot}. Since \cite{jacobson2014athletics} argued that executive functions can be improved by active participation in sports, the generalisation of existing results to an athlete population is therefore limited. Nevertheless, general studies conducted so far establish that executive functions are attributed to the frontal lobes of the brain \citep{stuss2000executive} which undergo protracted maturation from childhood to early adulthood \citep{lebel2008microstructural, taylor2013typical}, and this is followed by cognitive declines \citep{dempster1992rise, jurado2007elusive} due to the attrition of dendrites during the ageing process. These developments give rise to an inverted U-shaped executive functions trajectory across the lifespan of an individual \citep{kail1994processing,cepeda2001changes,zelazo2004executive}. \cite{huizinga2006age} and \cite{zelazo2011executive} documented that improvements in executive functions occur the most rapidly from late childhood into adolescence. In fact, \cite{diamond2002normal} reported that children between the age of 12 and 15 years old attain adult levels of performance in neuropsychological assessments.}

\textcolor{black}{This paper analyses the age-related architecture of executive functions in a sample of male elite youth soccer players from a professional German club in the Bundesliga, and compares our findings with existing theories. A gender- and sport-specific study is necessary here as evidence has shown that the impact of athletic participation varies significantly between gender \citep{habacha2014effects} and sport type \citep{krenn2018sport}. Particular emphasis is given to examining the developmental changes in executive function trajectories between late childhood (10--12 years old) and early adulthood (18--21 years old) since these periods are of relevance to talent identification and development. Contrary to previous investigations in the sport science literature which are based on cross-sectional data \citep[see e.g.][]{verburgh2014executive, huijgen2015cognitive, sakamoto2018possible}, our analysis uses longitudinal cognitive data collected from a test battery of soccer and non-soccer related neuropsychological assessments performed by the players over a period of three years. These outcomes are modelled using a latent variable model \citep{dunson2000bayesian, muthen2002beyond, proust2006nonlinear}, which relates speed and accuracy observables to latent variables representing executive functions. Furthermore, we make a distinction between domain-generic and domain-specific executive functions, in accordance with the two-component intellectual development model introduced by \cite{li2004transformations}. Indeed, \cite{furley2016working} recommended longitudinal studies that assess both domain-generic and domain-specific executive functions jointly, rather than examining them independently. This is done by modelling parameters of the underlying latent variables using a multivariate formulation.}

The rest of the paper is organised as follows. Section~\ref{sec:description} describes the data and the neuropsychological assessments performed by the players. Section~\ref{sec:methods sports} describes the latent variable model used in our analysis.  Section~\ref{sec:results} presents the results, and Section~\ref{sec:conclusion sports} concludes.

\section{Background of study} \label{sec:description}

\begin{figure}
\centering
\begin{subfigure}[t]{.35\textwidth}
  \centering
  \includegraphics[trim=0cm 0cm 0cm 12cm,clip,height=5.5cm]{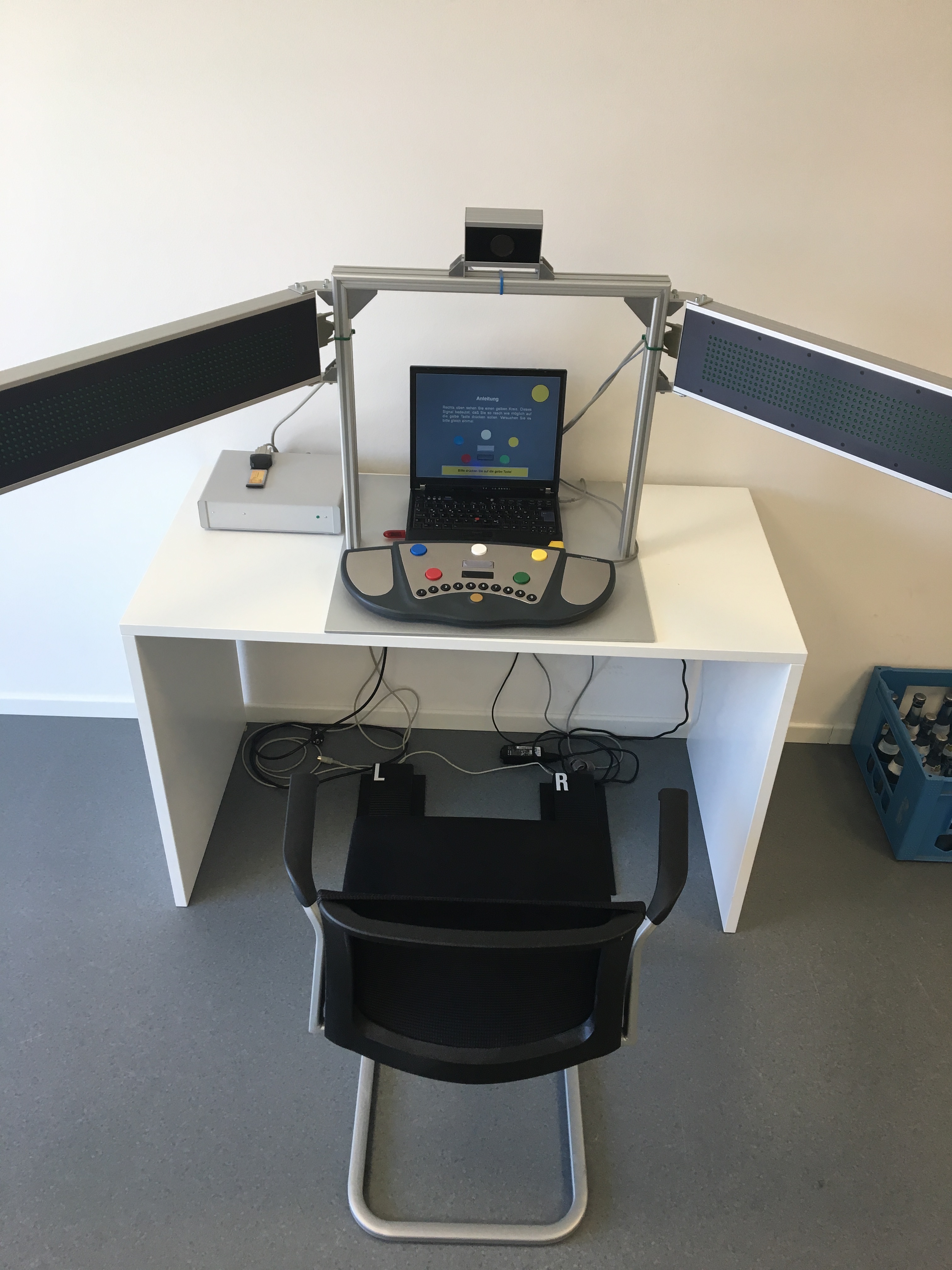}
  \caption{Determination test.}
\label{fig:determination}
\end{subfigure}
\hfill
\begin{subfigure}[t]{.62\textwidth}
  \centering
  \includegraphics[trim=0cm 0cm 0cm 0cm,clip,height=5cm]{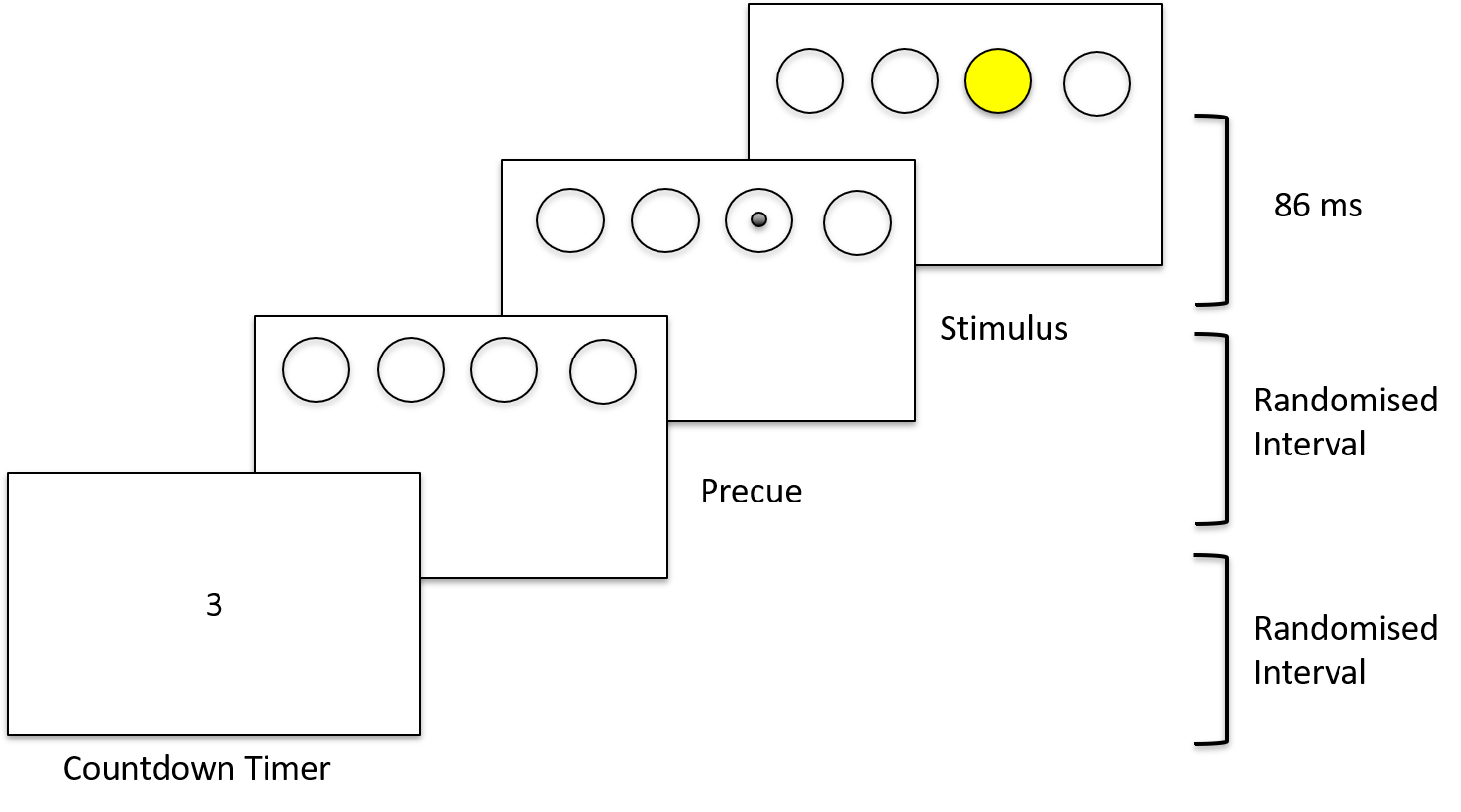}
  \caption{Pre-cued choice response time task.}
\label{fig:choice reaction}
\end{subfigure}
\vskip16pt
\begin{subfigure}[t]{.48\textwidth}
  \centering
  \includegraphics[trim=4cm 0cm 0cm 0cm,clip,width=\textwidth]{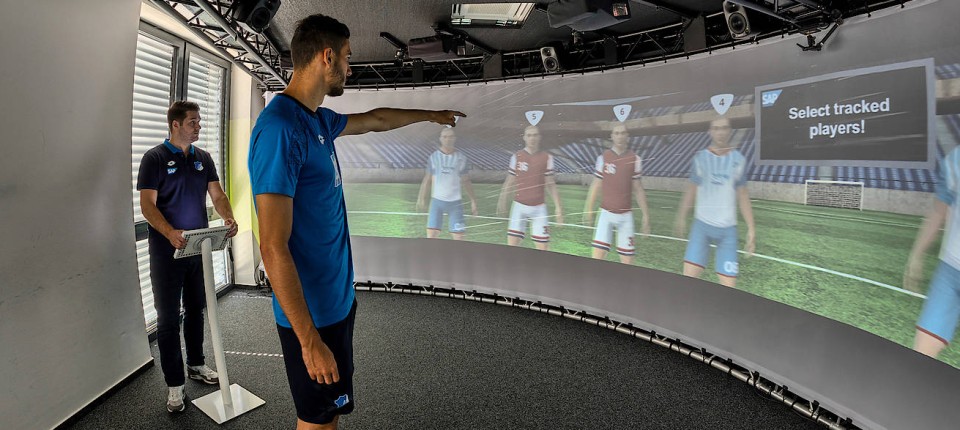}
  \caption{Helix test.}
\label{fig:helix}
\end{subfigure}
\hfill
\begin{subfigure}[t]{.49\textwidth}
  \centering
  \includegraphics[trim=3cm 1.5cm 1.5cm 0cm,clip,height=7cm]{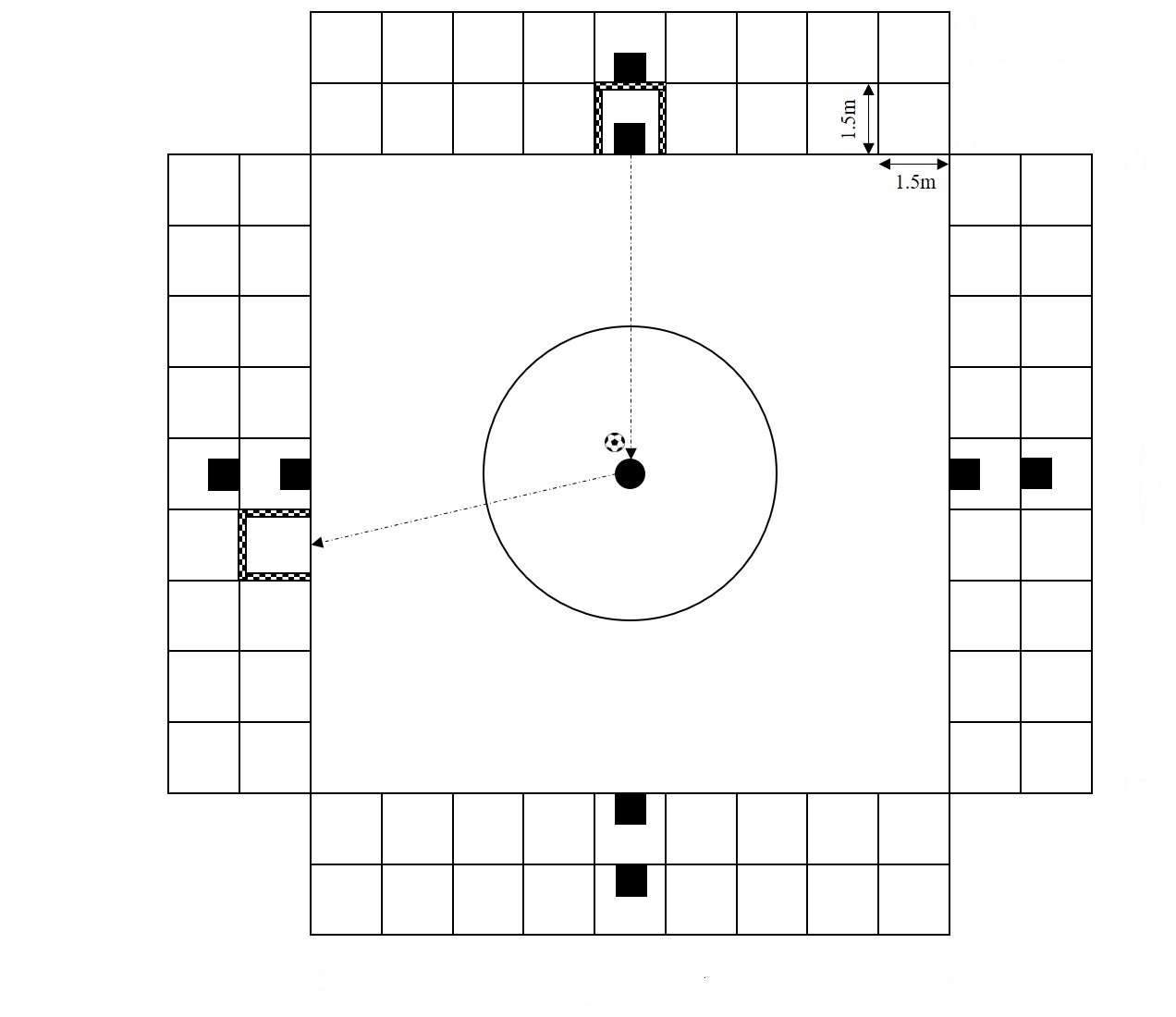}
  \caption{Footbonaut.}
\label{fig:footbonaut}
\end{subfigure}
\caption{\small Graphical illustrations for some of the neuropsychological assessments used in the study. (a) The test equipment used for the determination test in which a participant is required to respond to different types of stimuli by pressing the appropriate buttons on the keyboard panel and foot pedal. The response inhibition test uses the same equipment, but with a simpler keyboard design and without the foot pedal. (b) A congruent trial (the pre-cue appears in the same circle as the stimulus) in the pre-cued choice reaction time task. (c) A participant in the midst of identifying the players whom he is assigned to track in the Helix test while being monitored by a staff member. (d) A design plan of the Footbonaut in which a participant (the solid circle) is required to pass the soccer ball from one of the dispenser gates (square panels with a solid square within) to the illuminated target gate (an empty square panel partially surrounded by a striped band.)}
\label{fig:neuropsychological}
\end{figure}

The study collects the outcome variables from a test battery of neuropsychological assessments undergone by elite soccer players. The assessments measuring executive functions used in the study include a determination test \textcolor{black}{to measure general perceptual abilities}, a response inhibition test \textcolor{black}{to measure motor impulsivity}, a pre-cued choice response time task \textcolor{black}{to measure vigilance under interference}, a Helix test \textcolor{black}{to measure soccer-specific perceptual abilities} and a Footbonaut test \textcolor{black}{to measure soccer-specific technical abilities}. Figure~\ref{fig:neuropsychological} illustrates these assessments, while the following subsections give further details on each assessment in turn.

\subsection{Determination test}

The determination test (Schuhfried GmbH, M\"{o}dling, Austria) is a multi-stimuli assessment used in sports such as motor racing \citep{baur2006reactivity} to analyse the perceptual-motor abilities of a participant.  The participant is presented with a combination of two audio tones (2\,000 Hz and 100 Hz) and five coloured signals on a computer screen, to which the participant must react by choosing the appropriate buttons on the keyboard panels via hand and foot responses.  The number of correct answers to the stimuli within the four-minute assessment session and the median response time based on correct responses are recorded.  Since the participants are required to respond correctly to as many stimuli as possible within the stipulated time, the assessment provides a good evaluation of the reactive stress tolerance and receptivity of the participants. \textcolor{black}{The validity and reliability of the determination test in measuring executive functions have been confirmed by several studies \citep{whiteside2003review,ljac2012toward,beavan2019age}.}

\subsection{Response inhibition test}

The response inhibition test (Schuhfried GmbH, M\"{o}dling, Austria) uses a stop signal paradigm, \textcolor{black}{and has been widely established to provide a good assessment of impulse control \citep[see e.g.][]{alderson2007attention,zhong2014functional,cox2016driving}}.  The assessment is made up of 100 trials, with each presenting a left- or right-pointing arrow to which the participant must respond by pressing the corresponding button on the keyboard panel.  Each signal is displayed on the computer screen for a period of one second, followed by a one-second lapse before the subsequent signal appears.  There are two types of signals: go signal (around three quarters of the total trials) which requires instant reaction from the participant, and stop signal (usually appears after a go signal) to which the participant must refrain themselves from responding.  The stop signal is indicated by a tone at a pitch of 1\,000 Hz for 100 milliseconds.  The difficulty of the assessment is adaptive, i.e. a correct response to a stop signal will increase the visual-audio delay between the appearance of the arrow and the tone in the next stop trial by 50 milliseconds (up to a maximum of 350 milliseconds), and vice versa.  The variable of interest that reflects the ability to override prepotent actions \citep{logan1994ability} is the stop signal reaction time (SSRT), which is calculated by deducting the mean stop signal delay from the mean reaction time.  Measurements recorded are the SSRT, the mean reaction time and the number of correct responses.

\subsection{Pre-cued choice response time task}

In a pre-cued choice response time task, the participant is required to press the correct button on a joystick panel as fast as possible in response to a visual stimulus.  Four blank circles are arranged side by side and presented on the screen after a three-second countdown timer is shown.  One of them turns yellow after a randomised interval of between 2 to 4 seconds.  Prior to the colour change, a small dot appears in the center of one circle.  A total of 24 trials are conducted in the assessment, half of which have the dot in the same circle that turns yellow (congruent trials; see Figure~\ref{fig:choice reaction}), and the rest in a different circle (incongruent trials).  The mean response time for correct answers is recorded. \textcolor{black}{This task was previously used to assess psycho-motor vigilance under interference in both general population \citep{barela2019age} and professional athletes \citep{beavan2019age}, thereby validating its use as a measure of executive functions.}

\subsection{Helix test}

The Helix test (SAP SE, Walldorf, Germany) is designed to train various aspects of a participant's perceptual abilities including sustained attention, decision-making, multiple object tracking, and peripheral vision using a soccer-relevant stimulus.  A participant stands facing a 180-degree curved screen where a soccer stadium is reproduced.  Eight soccer players, with equal representation from two teams, are depicted as human-size avatars differing in the jersey colour, but otherwise having the same physical features and jersey number.  During the assessment, the animated avatars run randomly across the virtual pitch away from the participant for eight seconds before lining up across the screen in a different order to the initial line-up.  The participant must then identify four of the avatars that he is assigned to track before the start of the trial (see Figure~\ref{fig:helix}).  Ten trials are conducted and each correct identification earns a point to a maximum score of 40 points. \textcolor{black}{Although the Helix test is a newly developed assessment tool, \cite{beavan2020rise} found that it distinguishes executive functions of professional soccer players based on their playing experience.}

\subsection{Footbonaut test}

The Footbonaut (CGoal GmbH, Berlin, Germany) is an innovative tool which aims to measure a participant's soccer-specific skill performance such as dribbling, passing and shooting \citep{beavan2019footbonaut}, as well as perceptual-motor abilities.  The assessment system is made up of a square artificial turf surrounded by four walls.  Each wall contains 18 square panels arranged side by side in two rows, where the two panels in the middle serve as the ball dispenser gates while the rest are the target gates.  All gates measure 1.5 meters $\times$ 1.5 meters in dimension and are fitted with light barriers and light-emitting diodes (LEDs).  During the assessment, a ball is dispensed from one of the eight possible dispenser gates at a speed of 50 kilometers per hour.  Immediately before the ball is dispensed, the LEDs along the perimeter of the gate light up and an audio signal is given to the participant.  This is followed by the same stimuli 0.8 seconds later from the target gate, to which the participant is required to pass the ball.  Thirty-two trials are conducted and the mean reaction time \textcolor{black}{for successful passes that enter the target gate} is measured using the light barriers. \textcolor{black}{\cite{saal2018reliability} investigated the validity of the passing test in the Footbonaut, and concluded that it offers a reliable method to differentiate between skilled and less-skilled soccer players.}

\subsection{Description of data} \label{subsec:data description}

Table~\ref{tab:data summary} summarises the measurement variables for each of the neuropsychological assessments that \textcolor{black}{were} collected on 304 male soccer players, aged between 10 and 21 years old, who \textcolor{black}{represented} a professional German club in the Bundesliga.
%who participate in the neuropsychological test battery. 
Repeated measurements on the players \textcolor{black}{were} recorded over a study period of three years from the 2016--17 season to the 2018--19 season, whereby an assessment session \textcolor{black}{was} conducted twice in a year -- pre-season (between July and August) and post-season (between January and February). The number of assessment sessions that an individual \textcolor{black}{participated} in \textcolor{black}{varied}, due to player mobility between soccer clubs through the transfer market, or player dropout from the soccer academy. Additionally, the pre-cued choice response time task \textcolor{black}{was} only integrated into the test battery from the start of the 2017--18 season, whereas the Helix test \textcolor{black}{was} excluded from the test battery throughout the entire 2018--19 season. These changes in the test battery setup \textcolor{black}{resulted} in an increased proportion of missing data, as well as a reduction in the mean number of observations per player for measurements under both of these assessments. More missing data \textcolor{black}{were} also present in the measured outcomes as a result of data mismanagement.
\begin{table}[t!]
    \centering
    \onehalfspacing
    \resizebox{\textwidth}{!}{
    \begin{tabular}{|c|c|c|c|c|}
    \hline
        Neuropsychological & \multirow{2}{*}{Variable} & \multirow{2}{*}{Outcome} & Mean number of & Proportion of \\
        assessment & & & observations per player & missing observations$^\ast$ \\ \hline
        \multirow{2}{*}{Determination} & $y_1$ & Number of correct answers & 2.98 & 0.03 \\
        & $y_2$ & log median response time & 2.98 & 0.03 \\ \hline
        \multirow{3}{*}{Response inhibition} & $y_3$ & log SSRT & 2.98 & 0.03 \\
        & $y_4$ & log mean response time & 2.87 & 0.07 \\
        & $y_5$ & Number of correct answers & 2.98 & 0.03 \\ \hline
        \multirow{2}{*}{Choice response} & $y_6$ & log mean response time (congruent) & 1.61 & 0.48 \\
        & $y_7$ & log mean response time (incongruent) & 1.61 & 0.48 \\ \hline
        Helix & $y_8$ & Number of correct answers & 1.73 & 0.44 \\ \hline
        \multirow{2}{*}{Footbonaut} & $y_9$ & Number of correct answers & 2.39 & 0.22 \\
        & $y_{10}$ & log mean response time & 2.39 & 0.22 \\ \hline
        \multicolumn{5}{l}{\footnotesize{$^\ast$ Complete data have one observation per participated assessment session.}}
    \end{tabular}}
    \caption{\small The mean number of observations per player and the proportion of missing observations for each outcome variable of the neuropsychological assessments in the executive functions test battery.}
    \label{tab:data summary}
\end{table}

\begin{figure}[t!]
    \centering
    \includegraphics[trim=1.5cm 1.5cm .25cm 1.25cm,clip,height=0.4\textheight]{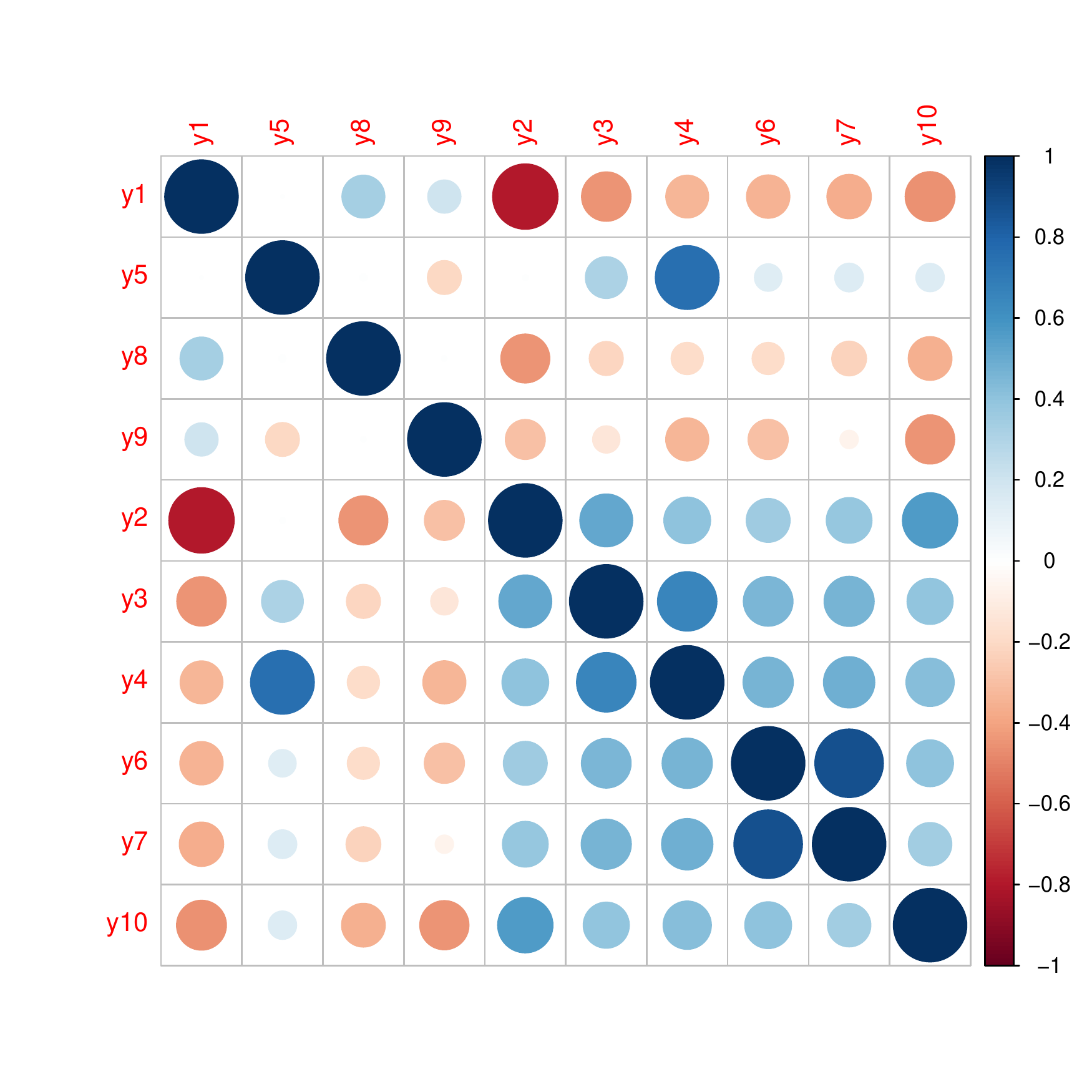}
    \caption{\small Spearman's rho correlation coefficients between the measurement variables collected from the 2017--18 pre-season assessment session. Circle size is proportional to correlation magnitude, with darker blue/red indicating stronger positive/negative correlation. Variables are ordered such that the first four $(y_1, y_5, y_8, y_9)$ report accuracy components  while the remainder $(y_2, y_3, y_4, y_6, y_7, y_{10})$ report speed components.}
    \label{fig:bivariate correlation}
\end{figure}

\textcolor{black}{Table~\ref{tab:data summary} shows the variables collected, which measure players' performance in accuracy and speed. Figure~\ref{fig:bivariate correlation} illustrates the pairwise dependence structure between these variables by computing their Spearman's rho correlation coefficients using complete observations from the 2017--18 pre-season assessment session (the only test battery that contains all five assessments). Although the assessments are very dissimilar in terms of their design and domain of cognitive abilities evaluated, the speed components $(y_2, y_3, y_4, y_6, y_7, y_{10})$ are noticeably strongly positively correlated among themselves. This suggests that players' speed is measured relatively consistently across different assessments. In contrast, the negative dependence observed in the speed-accuracy pairs from the determination test $(y_1, y_2)$ and the Footbonaut test $(y_9, y_{10})$ implies that higher-scoring players also tend to have shorter response time. This trend is in contrast to the speed-accuracy trade-off (longer movement time for increased accuracy element of a task) documented among soccer players in the experimental study conducted in \cite{andersen2011influence}.}

\begin{figure}[t!]
    \centering
    \includegraphics[trim=3cm 0.75cm 3cm 1.75cm,clip,height=0.4\textheight]{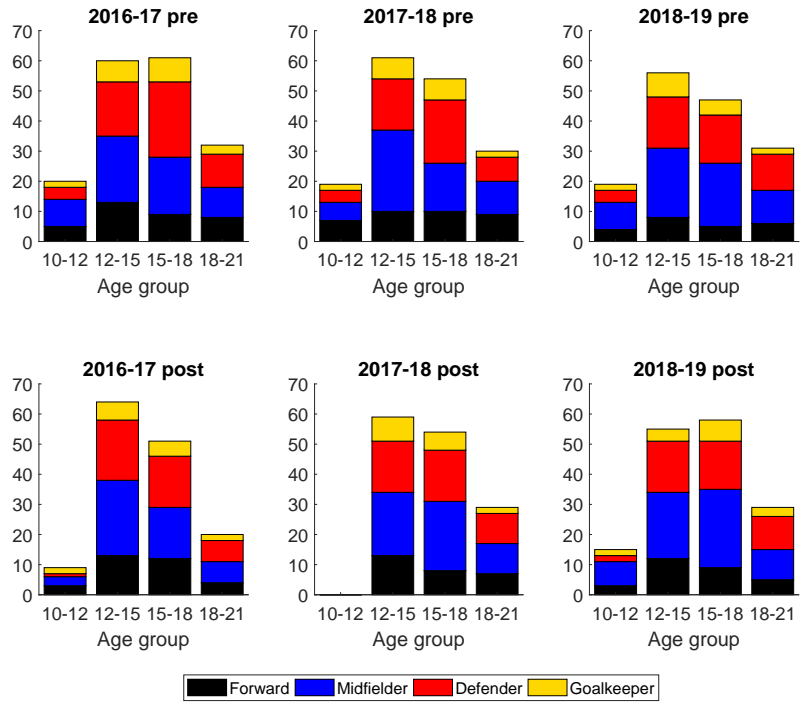}
    \caption{\small Bar charts showing the distribution of players by age group and playing position across the 3-year, pre- and post-season study period.}
    \label{fig:position age group}
\end{figure}

\textcolor{black}{Each participating player can be grouped according to their playing position in the team, either as forward, midfielder, defender or goalkeeper. According to the developmental model of sport participation in \cite{cote2007practice}, players' age range can be divided into four age groups that represent different developmental stages: 10--12 years old (late childhood), 12--15 years old (pre-adolescence), 15--18 years old (adolescence) and 18--21 years old (early adulthood). Figure~\ref{fig:position age group} shows the distribution of players by age group and playing position in each assessment session.
%A visual inspection of the bar charts indicates that 
More than two-thirds of players are aged between 12 and 18 years old when they participate in the test battery, whereas there are only at most 20 players in the youngest age group (10--12 years old). In terms of playing position, most players are either midfielders or defenders. Unsurprisingly, the goalkeeper category has the fewest number of players due to the composition of a soccer team which comprises ten field players and only one goalkeeper.}

\begin{figure}[t!]
    \begin{subfigure}[b]{0.485\textwidth}
        \centering
        \includegraphics[trim=1.5cm 1cm 1.5cm 1.25cm,clip,width=.9\textwidth]{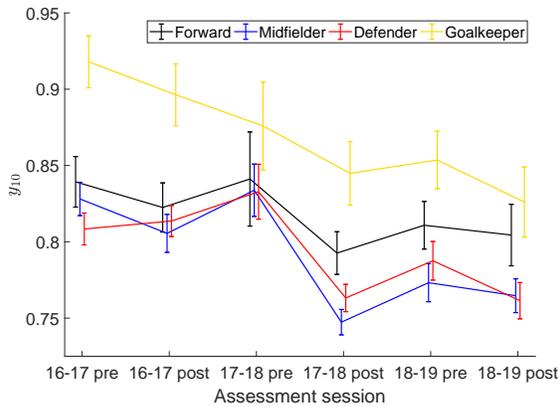}
        \caption{Changes in the grouped mean log response time measurement from the Footbonaut test across the study period. Error bars indicate two standard errors of the mean.}
        \label{fig:position pre post effect}
    \end{subfigure}
    \hfill
    \begin{subfigure}[b]{0.485\textwidth}
        \centering
        \includegraphics[trim=1.5cm 0cm 1.5cm 1.25cm,clip,width=.9\textwidth]{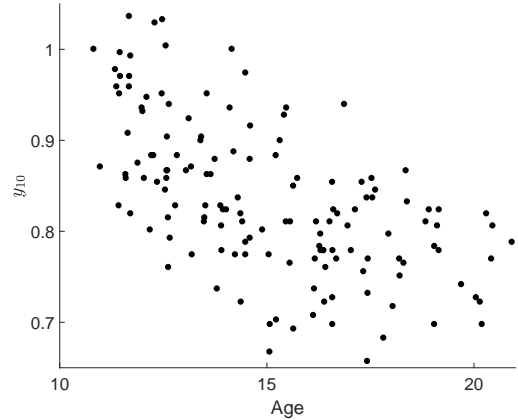}
        \caption{Scatterplot showing the relationship between the log response time measurement from the Footbonaut test and the player age, in the 2016--17 pre-season assessment session.}
        \label{fig:footbonaut scatterplot}
    \end{subfigure}
    \caption{\small Exploratory data analysis to examine performance variation between players that is due to assessment session, playing position and age.}
    \label{fig:exploratory}
\end{figure}

\textcolor{black}{Using the covariates available (assessment session and playing position), we 
perform an exploratory analysis to examine whether there are systematic differences in cognitive abilities between different groups of players. We demonstrate the general underlying pattern with the example of log response time measurement $y_{10}$ from the Footbonaut test. Figure~\ref{fig:position pre post effect} shows the changes in mean value of $y_{10}$ grouped by playing position across the study period. The goalkeepers consistently improve on their response time in the first four sessions, whereas the other field players' performance fluctuates considerably. Goalkeepers also take the longest time to react to the stimuli in the Footbonaut test. 
%Preliminary analysis result suggests that all 
Visually, and on average, players react faster in post-season assessment sessions within the same season, with a possible exception of defenders in the 2016--17 season. This improvement in post-season cognitive functioning can be attributed to the effect of active athletic participation \citep{jacobson2014athletics}. Overall, there is a general improvement in the response time as the study progresses. However, it must be recognised that this positive shift in performance level is likely to be confounded by the age of the players.}

%In order to investigate the age effect, 
\textcolor{black}{Figure~\ref{fig:footbonaut scatterplot} explores this age effect by showing differences in log response time as a function of age, based on observations from the 2016--17 pre-season assessment session (observations from other sessions produced similar graphs). The clear negative correlation 
%shown in the scatterplot serves as an evidence 
underlines that performance heterogeneity between players is age-dependent. Response time decreases significantly between ages 10 and 15, with a reduced rate of improvement as players age further. The age effect also offers possible explanation of two anomalies observed in Figure~\ref{fig:position pre post effect}: (i) post-season deterioration in the defenders performance in the 2016--17 season and, (ii) a sharp reduction in post-season response time for all players in the 2017--18 season. For the former, the session effect is offset because there are fewer defenders in the older 15--18 and 18--21 age groups (see Figure~\ref{fig:position age group}) who tend to perform better and are able to improve the group mean. For the latter, the age effect is amplified as there is no player below 12 years old (see Figure~\ref{fig:position age group}), resulting in the mean response time values being significantly lower.}

\section{Methods} \label{sec:methods sports}

Formal statistical modelling of the evolution of the players' executive functions requires consideration of the longitudinal nature of the study, as well as the fact that cognitive performance is determined by a battery of different but correlated neuropsychological assessments. Analysis of test outcomes is typically carried out using a multivariate linear mixed effects model \citep[e.g.][]{hall2001estimation, sliwinski2003correlated} whereby a unique mean latent growth trajectory that is representative of the entire population is estimated for each measured outcome. However, \cite{salthouse1996interrelations} establish in an experimental study that age-related differences in~the observed psychometric outcomes are not exclusively attributable to assessment-specific cognitive processes, but instead are the manifestation of a common cognitive factor. Furthermore, the two-component characterisation of lifespan intellectual development \citep{li2004transformations} distinguishes between two distinct facets of cognitive processes, namely fluid and crystallised abilities. The fluid abilities refer to biological and genetically pre-disposed intelligence in information processing, whereas the crystallised abilities relate to the normative and pragmatic aspects of expertise acquired through contextualised personal experiences and socio-cultural influences.

In the current context, the determination test, the response inhibition test and the pre-cued choice response time task examine general intelligence, and thus can be categorised as assessments that measure fluid (domain-generic) executive functions. In contrast, the Helix and the Footbonaut are specialised test systems designed to evaluate soccer-related skills that are developed through active participation in the sport, thereby demonstrating their roles in measuring crystallised (domain-specific) executive functions. Using this two-class grouping of neuropsychological assessments, we introduce two latent curves, one shared between domain-generic variables $(y_1, \dotsc, y_7)$ and another shared between domain-specific variables $(y_8, y_9, y_{10})$, within a latent variable model \citep{dunson2000bayesian, muthen2002beyond, proust2006nonlinear}. In the following we describe a latent process representing age-related changes in executive functions in a structural model. This model is then related to recorded outcomes of the corresponding neuropsychological assessments through a measurement model.

\subsection{The structural model}
\label{sec:3.1}

We model the latent curve underlying either of the two facets of executive functions (domain-specific and domain-generic) according to a piecewise linear spline model, which is a commonly used method in the epidemiological literature  to model longitudinal growth curves \textcolor{black}{(e.g.~\citealp{werner2006growth,de2011identifying,anderson2018comparing,chin2019multiclass})}. In the current setting, this models $\zeta_i \in \mathbb{R}$, the unobserved executive function for player $i$ at age $\omega_{it}$, as
\begin{equation}
    \zeta_i(\omega_{it}) = \b_{0i} (\omega_{it} - (\omega_{it} - \xi_1)_+) + \b_{\textcolor{black}{K}i} (\omega_{it} - \xi_{\textcolor{black}{K}})_+ + \sum_{\textcolor{black}{k}=1}^{\textcolor{black}{K}-1} \b_{\textcolor{black}{k}i} ((\omega_{it} - \xi_{\textcolor{black}{k}})_+ - (\omega_{it} - \xi_{\textcolor{black}{k+1}})_+), \label{eqn:broken stick sports}
\vspace{-.2em}
\end{equation}
\begin{equation}
    \bb_i = (\b_{0i}, \dotsc, \b_{\textcolor{black}{K}i})^\top \sim \N(\bmu_{\bb}, \Sig_{\bb}), \label{eqn:random effects sports}
\end{equation}
for $i=1,\dotsc,304$, where $(a)_+ = \max\{0, a\}$ is the positive part of $a$, and $\bxi = (\xi_1, \dotsc, \xi_{\textcolor{black}{K}})^\top$ is an ordered vector of $\textcolor{black}{K}$ knots such that $\xi_1 < \dotsc< \xi_{\textcolor{black}{K}}$. 
\textcolor{black}{Here, $t$ measures time on the scale of the age of each individual, so that e.g.~$\zeta_i(\omega_{it})$ with $\omega_{it}=10$ refers to the executive function of individual $i$ when the individual is 10 years old.}
In this way, (\ref{eqn:broken stick sports}) models the evolution of a player's executive functions over time \textcolor{black}{(age)} by $\textcolor{black}{K+1}$ piecewise linear segments with breaks at $\bxi$. Greater flexibility can be achieved by introducing player-specific random change points, but here we fix $\bxi$ due to the sparsity of the data. The vector of random slope coefficients $\bb_i = (\b_{0i}, \dotsc, \b_{\textcolor{black}{K}i})^\top$, where $\b_{\textcolor{black}{k}i}$ represents the rate of change in $\zeta_i$ between knots $\xi_{\textcolor{black}{k}}$ and $\xi_{\textcolor{black}{k+1}}$, allows for the heterogeneity between players that is due to unobservables such as the number of training hours \textcolor{black}{and familiarity with the test}.  By assuming a Gaussian distribution on the vector of random effects $\bb_i$ in (\ref{eqn:random effects sports}), we postulate that the mean rate of developmental growth in executive functions in the population is $\bmu_{\bb}$, and that the variability of player-specific deviations from this global trend is characterised by $\Sig_{\bb}$.

\subsection{The measurement model}

Write $y_{d,it}$ as the observed test outcome $d$ for player $i$, on the $t$-th measurement occasion, where $i=1, \dotsc, 304,$ $t=1, \dotsc, T_i$ and $d=1, \dotsc,10$.
Assume, for the moment, that each $y_{d,it}$ is a continuous measurement (this is relaxed below). Using a latent variable model \citep[e.g.][]{dunson2000bayesian, muthen2002beyond, proust2006nonlinear}, we link the unobserved executive functions to the outcome variable by
\begin{equation}
    y_{d,it} = \a_d + \bx_{it}^\top \bm{\g}_d + c_{d\ell} \zeta_{i\ell}(\omega_{it})  + {\eps}_{d,it}, \quad \bm{\eps}_{it} = (\eps_{1,it}, \dotsc, \eps_{D,it})^\top \sim \N(\bm{0}, \Sig_{\bm{\eps}}),
    \label{eqn:measurement 1}
\end{equation}
where $\a_d$ is an intercept for test outcome $d$ and $\bx_{it}$ is a vector of time-dependent player-specific covariates associated with a vector of fixed effects $\bm{\g}_d$ for test $d$. 
\textcolor{black}{The recorded covariates for this study include a player's playing position (forward, midfield, defence or goalkeeper) and an indicator for post-season assessment measurements.}
In this way, the covariates can vary in their effects on each test outcome $d$.   The error term $\eps_{d,it}$ is assumed to be uncorrelated with the exogenous variables $\bx_{it}$ and the latent executive functions $\zeta_i$. To incorporate the distinction between the two facets of cognitive processes, we introduce an additional index $\ell \in \{1,2\}$ on the executive functions in (\ref{eqn:measurement 1}) such that $\ell=1$ when $d=1, \dotsc, 7$ refers to domain-generic functions, and $\ell=2$ when $d=8,\ldots,10$ refers to domain-specific functions.
\textcolor{black}{As a result, the term $\a_d+c_{d\ell} \zeta_{i\ell}(\omega_{it})$ stipulates a test-level linear rescaling (to account for scale differences in the outcome variables) of either the domain-generic or domain-specific executive functions, which is itself an individual-specific random effect around a population mean as detailed in Section \ref{sec:3.1}.}
%
%The coefficient $c_{kl}$ adjusts for the scale difference in the outcome variables. 
In order for all model parameters to be identifiable, one of the $c_{d\ell}=1$ for each value of $\ell$, i.e.~\textcolor{black}{$c_{d1}=1$ for one $d \in \{1, \dotsc, 7\}$ and $c_{d2}=1$ for one $d \in \{8, \dotsc, 10\}$, and $e_{it}$ is restricted to a standard normal distribution $\N(0,1)$. We choose the value of $d$ for which $c_{d\ell}=1$ to be the measurement with the largest scale so that the magnitude of $c_{d\ell}$ is less than 1 for other measurements.}

Some of the measured outcomes are count variables, e.g.~the number of correct answers in the determination test. As a result, the assumption of normality on the errors for these outcomes in (\ref{eqn:measurement 1}) may be unsuitable. To account for this, we follow \cite{gelman2013bayesian} and transform the count variables into continuous outcomes using the Gaussian kernel. In particular,
\begin{equation*}
    y_{d,it} = h(y_{d,it}^\ast), \quad d \in \{1,5,8,9\},
    \label{eqn:poisson}
\end{equation*}
where $h(\cdot)$ is a rounding function such that $h(y^\ast) = p$ if $y^\ast \in (a_p, a_{p+1}]$ for $p=0, \ldots, \infty$, $a_0=-\infty$, and $a_p=p-1,$ for $p=1, \ldots, \infty$. \textcolor{black}{The latent continuous variable $y^*_{d,it}$ is then modelled following the measurement model in \eqref{eqn:measurement 1}.} As mentioned in Section~\ref{subsec:data description}, the non-availability of certain neuropsychological assessments and \textcolor{black}{data management practices} have resulted in the outcome vectors $\by_{it}=(y_{1,it}, \dotsc, y_{D,it})^\top$ being partially observed. We overcome the missingness \textcolor{black}{by fitting the model assuming full data, and the missing values are sampled from their full conditional distributions (which are the posterior predictive distributions) in the Bayesian sampling scheme.}

\subsection{Prior distributions and implementation}
Posterior inference for the above model can be performed in the Bayesian framework, implemented via Markov chain Monte Carlo \citep{robert2004monte}. For prior distributions, a horseshoe prior \citep{makalic2016simple} is specified on $\bmu_{\bb}$ and $\bm{\g}_d,d=1,\dotsc,10$, \textcolor{black}{which is designed to have concentration at zero to shrink small coefficients towards zero, while having heavy tails to avoid over-shrinkage of larger coefficients}. A hierarchical inverse-Wishart prior \citep{huang2013simple} with 2 degrees of freedom and scale parameter 25 is chosen on $\Sig_{\bb}$ and $\Sig_{\bm{\eps}}$ \textcolor{black}{to induce a sparse structure on the partial correlation matrices \citep{chin2020efficient}}, while each $\a_d$ has a standard non-informative $\N(0, 10^3)$ prior.

For parameter identifiability, the scale coefficients for the number of correct answers in the determination test and the Helix test are set as one (i.e.~$c_{11} = c_{81} = 1$). \textcolor{black}{These are chosen as $y_1$ and $y_8$ have the largest scales among the variables measuring the two facets of executive functions, and so this ensures that each of the other $c_{d\ell}$ will typically scale around or less than 1 in magnitude. Assessment tasks associated with the same type of assessment group (speed or accuracy) are likely to be positively correlated, but those tasks are likely to be negatively correlated between these two groups. This is largely evident in Figure~\ref{fig:bivariate correlation}. Since both $y_1$ and $y_8$ are accuracy components,} an informative $\N(0.5, 0.25)$ prior is specified on $c_{d\ell}$ if measurement $d$ also relates to an accuracy component of the assessment $(y_5,y_9)$ to express the prior belief that $c_{d\ell}$ is likely to be a value between 0 and 1. Conversely, a $\N(-0.5,0.25)$ prior is used if the measurement relates to log speed $(y_2,y_3,y_4,y_6,y_7,y_{10})$.

\section{Analysis and results} \label{sec:results}

\begin{figure}[t!]
    \centering
    \includegraphics[width=\textwidth]{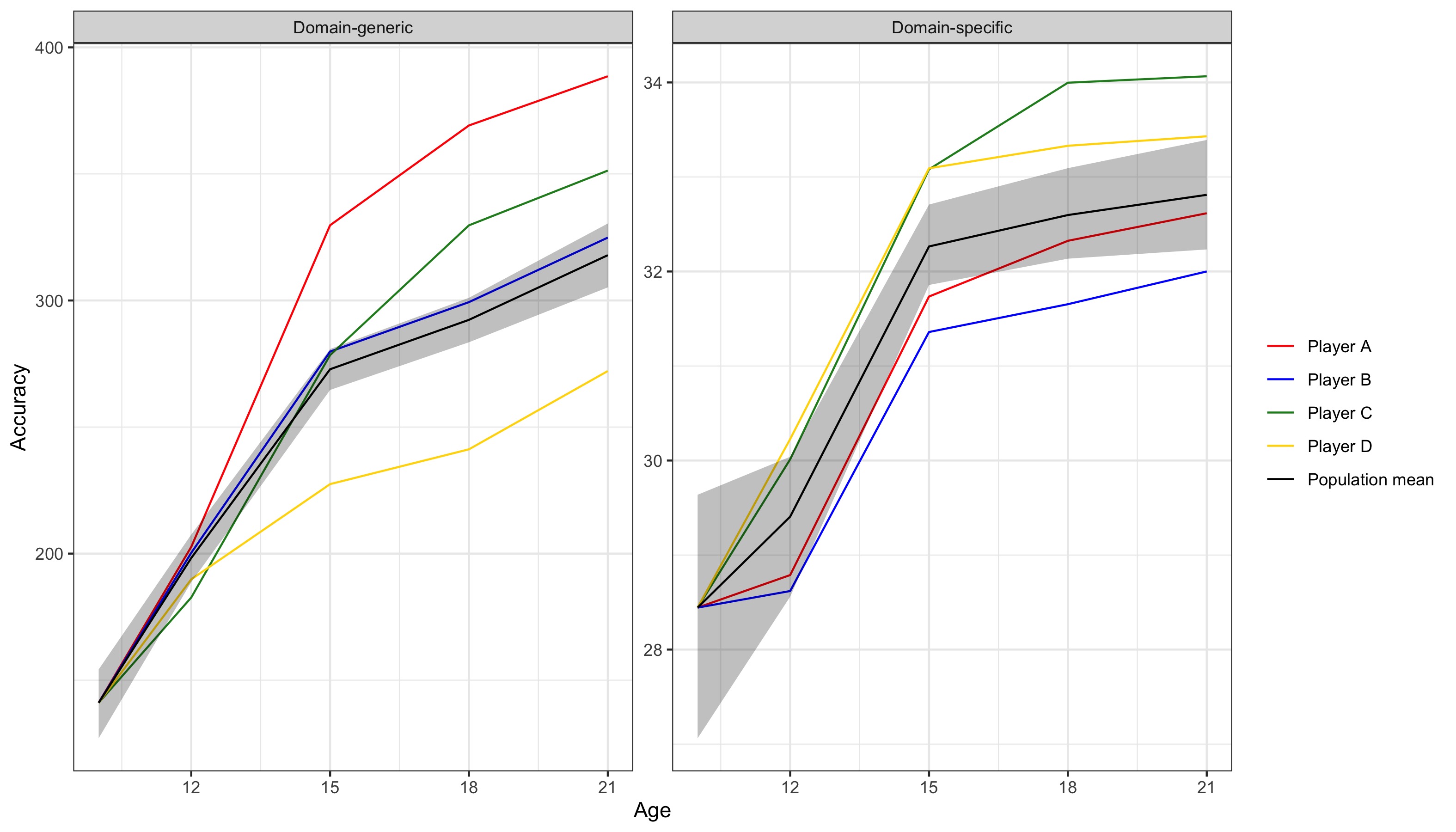}
    \caption{\textcolor{black}{\small Domain-generic (left) and domain-specific (right) executive functions trajectories for a sample of players (coloured lines) and the population mean (black lines)  of the population.
    %, based on the accuracy of the determination test and the Helix test respectively. 
    The grey shaded region indicates pointwise 95\% HPD credible intervals of the population mean trajectories.}}
    \label{fig:executive function}
\end{figure}

Following the age groupings defined in the context of sport \citep{cote2007practice}, we consider the development of executive functions in the four different stages of growth: late childhood (10--12 years old), pre-adolescence (12--15 years old), adolescence (15--18 years old) and early adulthood (18--21 years old).  Accordingly, the three knot locations ($\textcolor{black}{K}=3$) in the piecewise linear spline model in (\ref{eqn:broken stick sports}) are specified as $\bxi = (12,15,18)^\top$. \textcolor{black}{Figure~\ref{fig:executive function} shows the estimated mean trajectories of the domain-generic and domain-specific executive functions for a chosen sample of players (coloured lines), and compares them to the population mean (black lines) whose pointwise 95\% highest posterior density (HPD) credible regions are represented by the shaded region. 
% SCOTT: I commented out the next sentence as I don't understand it.
%For simplicity, these trajectories are based on the number of correct answers scored in the determination test (domain-generic) and the Helix test (domain-specific) whereby the scale factor $c_{d\ell}$ is assumed a fixed value of 1. 
The results indicate that changes in executive functions of the elite soccer player population occur mainly between 10 to 15 years of age since the magnitudes of the slopes within this age range are the largest. In particular, for domain-generic executive functions the most rapid increase happens during late childhood (10--12 years)  (a posterior mean rate of 28.59, which is marginally higher than the value of 24.86 in the next period), whereas for domain-specific executive functions the most rapid increase happens during pre-adolescence (12--15 years) (a posterior mean rate of 0.95, which is nearly twice as large as the value of 0.48 in late childhood). Domain-generic executive functions continue to develop, albeit at a much slower pace, during adolescence (an average rate of 6.51) and early adulthood (a posterior mean rate of 8.52). This observation is consistent with the findings in \cite{diamond2002normal}, which argues that performance in domain-generic executive functions reaches adult performance levels between 12 and 15 years of age.} 

\textcolor{black}{Given that maturation in domain-specific executive functions is largely conditioned by occupational expertise \citep{li2004transformations}, it could be hypothesised that improvements in soccer-related abilities will be reflected in the trajectory across all developmental stages as players are continuously challenged to refine their skills in order to remain competitive \citep{mann2007perceptual}. However, our results show that the increase in domain-specific executive functions is almost negligible after 15 years of age (with posterior mean rates of 0.11 and 0.07 during adolescence and early adulthood respectively). A possible explanation for the observed plateau is that both the domain-generic and domain-specific assessments used in our study do not necessarily represent} the way in which perception and action of competition are coupled in soccer \citep{pinder2011manipulating}. For example, while the Footbonaut test has some validity for measuring soccer skills, it requires players to use passing actions to react to visual and auditory stimuli that are unrelated to soccer \citep{beavan2019footbonaut}. \textcolor{black}{Meanwhile, the Helix test measures high-level perceptual abilities specific to soccer but its design lacks an action component. Therefore, it may be unsurprising that the expected positive association between domain-specific executive functions and age as a proxy for soccer experience is not observed.}

\textcolor{black}{When inspecting the posterior distribution of the dependence structures in $\Sig_{\bb}$ (results not shown), we find that the slopes of the piecewise linear spline model are independent of each other and across both facets of executive functions. This suggests that previous studies which are based on cross-sectional data \citep{verburgh2014executive, huijgen2015cognitive, vestberg2017core, sakamoto2018possible} may have overstated the usefulness of domain-generic executive functions in soccer talent identification. This is because a strong relationship between both types of executive functions should exist if domain-generic executive functions are a prognostic tool for soccer performance. A sample of four players were chosen and shown in Figure~\ref{fig:executive function} (coloured lines) to illustrate this argument. We observe that players A and B demonstrate an above-average level of domain-generic executive functions, but fail to reproduce similar level of superiority in soccer-specific assessments. Conversely, player D who has less developed domain-generic executive functions compared to the population mean outperforms his peers in terms of soccer expertise. Player D also achieve a comparable level of domain-specific executive functions to player C although the latter performs better in the generic abilities test battery.
}
\begin{table}[b!]
    \centering
    \onehalfspacing
    \begin{tabular}{|c|rrrrr|}
    \hline
        Variable & Intercept & Post-season & Forward & Midfielder & Defender \\ \hline
        $y_1$ & \cellcolor{gray!25}141.08 & 1.43 & 0.00 & $0.00$ & $-0.01$ \\
        $y_2$ & 0.01 & \cellcolor{gray!25}$-0.03$ & $0.00$ & 0.00 & $0.00$ \\
        $y_3$ & \cellcolor{gray!25}$-1.26$ & \cellcolor{gray!25}$-0.09$ & 0.00 & $-0.04$ & 0.00 \\
        $y_4$ & \cellcolor{gray!25}$-0.64$ & $-0.02$ & 0.00 & $0.00$ & $0.00$ \\
        $y_5$ & \cellcolor{gray!25}81.85 & $-0.22$ & 0.06 & 0.02 & $-0.34$ \\ 
        $y_6$ & \cellcolor{gray!25}$-0.49$ & \cellcolor{gray!25}$-0.04$ & 0.00 & 0.00 & 0.00 \\
        $y_7$ & \cellcolor{gray!25}$-0.44$ & \cellcolor{gray!25}$-0.04$ & $0.00$ & 0.00 & $0.00$ \\
        $y_8$ & \cellcolor{gray!25}28.44 & 0.01 & $-0.02$ & 0.02 & $0.00$ \\ 
        $y_9$ & \cellcolor{gray!25}22.93 & 0.15 & $-0.60$ & 0.23 & 0.27 \\
        $y_{10}$ & \cellcolor{gray!25}1.03 & $-0.01$ & \cellcolor{gray!25}$-0.05$ & \cellcolor{gray!25}$-0.06$ & \cellcolor{gray!25}$-0.04$ \\ \hline
    \end{tabular}
    \caption{\small Estimated posterior means of regression coefficients $\bm{\g}_d$ for the covariates for each outcome variable. Parameters whose 95\% HPD credible interval does not include 0 are highlighted in grey.}
    \label{tab:covariates}
\end{table}

\textcolor{black}{We now examine the impact of the covariates on each outcome variable in the test battery.
Table~\ref{tab:covariates} shows the regression coefficient posterior mean estimates for the assessment session and playing position.} The players tend to have better response times $(y_2, y_3, y_6, y_7)$ in domain-generic tasks during post-season assessment sessions, indicating that there is an acute effect of soccer participation on performance in these assessments. \textcolor{black}{The absence of positional effects in the Helix test further reinforces our previous argument that its design may not have adequately coupled perceptual information with soccer-specific actions.} We also observe that goalkeepers generally perform the worst in the Footbonaut test in terms of the response time $y_{10}$ (i.e.~players in the other positions respond much faster). This is because as part of their training, goalkeepers tend not to train receiving, control and passing of the ball to the same extent as players in other positions. As a result, we can conclude that the Footbonaut test represents a more useful measure of performance for field players rather than for goalkeepers.

\section{Conclusion} \label{sec:conclusion sports}

%Soccer is an example of an open skills sport \citep{wang2013open}, requiring players to react in a dynamically changing, unpredictable and externally-paced environment.  It has been shown in recent research \citep{vestberg2012executive, huijgen2015cognitive, sakamoto2018possible} that executive functions play a central role in predicting future performance of players, and hence provide useful information to professional clubs in talent identification programmes.  

\textcolor{black}{This paper explores the relationship between age and executive functions in a professional athlete population by modelling the cognitive outcomes from a test battery of neuropsychological assessments performed by elite soccer players in a longitudinal study using a latent variable model \citep{dunson2000bayesian,muthen2002beyond, proust2006nonlinear}. The findings of previous research on the developmental trajectories of executive functions were based on cross-sectional studies \citep{verburgh2014executive, huijgen2015cognitive, sakamoto2018possible}. To the best of our knowledge, this is the first study of its kind in the sport science literature that is based on longitudinal data. The latent growth curve representing the unobserved executive functions was modelled as a piecewise linear function across time within a random effects model, and was linked to the observed outcomes via a measurement model. Following the argument in \cite{li2004transformations}, we differentiated between fluid (domain-generic) and crystallised (domain-specific) executive functions, where the former develops biologically while the latter is acquired through occupational experience. This allowed us to make a comparison between their trajectories across different stages of growth development \citep{cote2007practice}. Rather than examining both types of executive functions independently, which is commonly done in the literature \citep{furley2016working}, we modelled them jointly in a multivariate formulation to investigate if the claims made on the importance of domain-generic executive functions as a prognostic tool for excellence in soccer can be substantiated.}

\textcolor{black}{
Our analysis showed that both facets of executive functions exhibit a rapid increase between 10 and 15 years of age, and while domain-generic executive functions continue to develop at a much slower rate, domain-specific executive functions begin to plateau after 15 years. The latter observation is contrary to the popular belief that soccer players who excel in competitive settings tend to possess more developed technical abilities shaped by their playing experience. However, the lack of evidence supporting this expectation in our study could possibly be due to the failure of the assessment design to reproduce the perception-action couplings experienced by players during an actual match \citep{pinder2011manipulating}. We also found no substantial dependencies in the rate of developmental growth between domain-generic and domain-specific executive functions, thereby contradicting the findings of earlier studies \citep{vestberg2012executive, verburgh2014executive, sakamoto2018possible} and weakening the argument that domain-generic executive functions provide useful information for soccer talent identification. The longitudinal nature of the study allowed our modelling approach to control for unobserved heterogeneity such as the number of training hours \citep{huijgen2015cognitive}, and hence providing a closer representation of the underlying mechanistic development in cognitive abilities. Based on our results, a clear conclusion is that the value of integrating neuropsychological test batteries in soccer talent identification programmes is debatable given that no interaction is established between domain-generic and domain-specific executive functions. A comprehensive study on the reliability of each neuropsychological assessment in the test battery should be undertaken to validate their use \citep{dicks2009representative}, especially if the results of these tests are used pervasively.
}

There are various possibilities to extend the model: the distributional assumption on the rate of change in the executive functions may be relaxed to a more flexible structured finite mixture distribution, or modelled non-parametrically using a Dirichlet process mixture model \citep{antoniak1974mixtures}. It is assumed in the model that the number of knots and their locations are known {\it a priori}, and thus a natural extension would consider inferring them as part of the Bayesian posterior sampling scheme by specifying a large number of equally spaced knots and placing spike and slab priors on the rate of change coefficients. However, these were not feasible with the current dataset, and would necessarily involve higher data collection costs.

\section*{Acknowledgements}

Vincent Chin, Scott A.~Sisson, Robert Kohn and Louise Ryan were partially supported by the Australian Research Council through the Australian
Centre of Excellence in Mathematical and Statistical Frontiers (ACEMS; CE140100049).

\bibliographystyle{chicago}
\interlinepenalty=10000
\bibliography{biblio}

\end{document}